\def \VersionAuthor {}
\documentclass[runningheads]{llncs}

\ifdefined\VersionAuthor
\else
	\makeatletter
	\AtBeginDocument{%
	\@ifpackageloaded{hyperref}
	{\def\@doi#1{\href{https://doi.org/#1}
		{\ttfamily https://doi.org/#1}\egroup}}
	{\def\@doi#1{\ttfamily https://doi.org/#1\egroup}}
	\def\doi{\bgroup\catcode`\_=12\relax\@doi}}
	\makeatother
\fi
\usepackage[ruled,vlined,linesnumbered]{algorithm2e}
	\SetKwInOut{Input}{input}
	\SetKwInOut{Output}{output}
\usepackage[utf8]{inputenc}
\usepackage[english]{babel}
\usepackage{booktabs}
\usepackage{csquotes}

 \usepackage{graphicx}
 \usepackage{amssymb,amsmath}%

\usepackage[misc,geometry]{ifsym} %

\ifdefined\VersionAuthor
	\usepackage[backend=biber,backref=true,style=alphabetic,url=false,doi=true,defernumbers=true,sorting=anyt,maxnames=99]{biblatex} %
	\renewbibmacro*{doi+eprint+url}{%
		\iftoggle{bbx:doi}
			{\color{black!40}\footnotesize\printfield{doi}}
			{}%
		\newunit\newblock
		\iftoggle{bbx:eprint}
			{\usebibmacro{eprint}}
			{}%
		\newunit\newblock
		\iftoggle{bbx:url}
			{\usebibmacro{url+urldate}}
			{}%
	}

\fi
\DeclareUnicodeCharacter{0301}{\'{e}}

\ifdefined\VersionWithComments
	\usepackage{draftwatermark}
	\SetWatermarkText{draft}
	\SetWatermarkScale{15}
	\SetWatermarkColor[gray]{0.9}
\fi
\usepackage[svgnames,table]{xcolor}
\definecolor{darkblue}{rgb}{0, 0, 0.7}

\usepackage[
		pdfauthor={Jawher Jerray and Laurent Fribourg},%
		pdftitle={Robust optimal control using dynamic programming and guaranteed Euler’s method},
		breaklinks  = true,
		colorlinks  = true,
	\ifdefined \VersionWithComments
		pagebackref = true,
	\fi
		citecolor   = blue!50!blue,
		linkcolor   = darkblue,
		urlcolor    = blue!50!green,
	]{hyperref}

\usepackage[capitalise,english,nameinlink]{cleveref} %
\crefname{line}{\text{line}}{\text{lines}} %

\usepackage{tikz}
\usetikzlibrary{decorations.pathmorphing}
\usetikzlibrary{arrows,calc}
\tikzset{
>=stealth',
help lines/.style={dashed, thick},
axis/.style={<->},
important line/.style={thick},
connection/.style={thick, dotted},
}

\ifdefined\VersionWithComments
	\usepackage[colorinlistoftodos,textsize=footnotesize]{todonotes}
\else
	\usepackage[disable]{todonotes}
\fi

\ifdefined \VersionWithComments
	\newcommand{\todoinline}[1]{\mbox{}{\color{red}{\textbf{TODO}\ifx#1\\\else:\ \fi #1}}} %
\else
	\newcommand{\todoinline}[1]{}
\fi

\usepackage{verbatim} %

\ifdefined \VersionWithComments
	\usepackage{soul}
	
	\newcommand{\reviewDelete}[1]{{\color{red}\st{#1}}}
\else
	
	\newcommand{\reviewDelete}[1]{}
\fi

\ifdefined\VersionAuthor
\else
	\usepackage[pagewise]{lineno} %
	\linenumbers
	
\fi

\ifdefined \VersionWithComments
 	\definecolor{colorok}{RGB}{80,80,150}
\else
	\definecolor{colorok}{RGB}{0,0,0}
\fi

\RequirePackage{amsmath}

\usepackage{mathabx}

\usepackage{color}

\makeatletter
\newcommand\footnoteref[1]{\protected@xdef\@thefnmark{\ref{#1}}\@footnotemark}
\makeatother

\author{%
Jawher Jerray\inst{1}
\and
Laurent Fribourg\inst{2}
\and
Étienne André\inst{3}\orcidID{0000-0001-8473-9555}
}

\institute{$^1$	Université Sorbonne Paris Nord, LIPN, CNRS, UMR 7030, F-93430, Villetaneuse, France \\
	$^2$ Université Paris-Saclay, LSV, CNRS, ENS Paris-Saclay\\
	$^3$ Université de Lorraine, CNRS, Inria, LORIA, F-54000 Nancy, France\\
  \email{jerray@lipn.univ-paris13.fr}
}

\graphicspath{{./figures/}}

\titlerunning{Robust optimal control using dynamic programming and Euler's method}

\authorrunning{J. Jerray, L. Fribourg, É. André}
\begin{document}
\title{Robust optimal control using dynamic programming and guaranteed Euler’s method}
\maketitle              %
\begin{abstract}
Set-based integration methods allow to prove properties of  differential systems, which take into account bounded disturbances.  The systems (either time-discrete, time-continuous or hybrid) satisfying such properties are said to be ``robust’’. In the context of optimal control synthesis, the set-based methods are generally extensions of numerical optimal methods of two classes: first,  methods based on convex optimization; second, methods based on the dynamic programming principle. Heymann et al. have recently shown that, for certain systems of low dimension, the second numerical method  can give better solutions than  the first one. They have built a solver (Bocop) that implements both numerical methods. We show in this paper that a set-based extension of a method of the second class 
which uses a guaranteed Euler integration method, allows us to find such
good solutions. Besides, these solutions enjoy the property of robustness against uncertainties on initial conditions and bounded disturbances.
We demonstrate the practical interest of our method 
on an example taken from the numerical Bocop solver.
We also give a variant of our method, inspired by the method of Model Predictive Control, that allows us to find more efficiently an optimal control
at the price of losing robustness. 

\end{abstract}

\section{Introduction}
Given a differential system with input of the form $\overset{.}{y}(t)=f(y(t),{\bf u}(t))$ and an initial condition $y(0)=y_0$, the calculation of a control ${\bf u}(\cdot)$  that minimizes a given cost function ({\em optimal control}) rarely has an analytical solution, and numerical methods must be used to obtain approximate solutions. Among these numerical methods, there are 3 main classes:
\begin{enumerate}
    \item methods that reduce the problem to a {\em convex optimization} problem, which are very much in demand since the adaptation of interior point methods by Nesterov and Nemirovskii \cite{NesterovN94}, and which are notably used in receding horizon methods, also called Model Predictive Control (MPC) \cite{Mayne14};
    \item methods based on the resolution of an Hamilton-Jacobi-Bellman (HJB) equation 
(see, e.g., \cite{Falcone1999,Saluzzi18}), using the {\em Dynamic Programming Principle} (DPP)~\cite{Bellman57};
    \item methods based on the {\em Pontryagin Maximum Principle} (PMP) \cite{kirk1970optimal}.
\end{enumerate}

On the other hand, since the 1960s and the invention of Interval Arithmetic~\cite{Moore66}, one has been looking for safe enclosures for the approximate values of ODEs computed by numerical methods. Therefore, extensions of numerical methods, called set-based (or symbolic) methods, have been used, which, instead of manipulating points, manipulate sets (typically real intervals or products of real intervals) in order to enclose the exact values. 
These methods of control synthesis are called ``correct-by-design'' or ``guaranteed''.
In addition to ensuring that a set (typically an interval) containing the exact solution is obtained at the end, set-based methods allow taking into account 
bounded disturbances.
They are said to be 
``robust''. 
Since the beginning of interval arithmetic, these set-based methods have experienced a great development. The manipulated sets, originally products of real intervals \cite{Moore66}, have taken specialized convex forms such as polytopes \cite{HanK06}, parallelotopes \cite{Lohner87}, zonotopes  \cite{girard2005reachability}, spheres \cite{SNR17} or ellipsoids~\cite{Neumaier1993}. In this context,
numerical integration methods classically take set-based forms using extensions of Taylor's methods (see, e.g., \cite{AlthoffSB07,BerzH98,BerzM98,CAS12,Lohner87,Nedialkov99,NedialkovKS04}). 

Numerical methods 1 and 2 of {\em optimal control} have themselves been subject to set-based extensions to take account of uncertainties (unlike method 3, which is very sensitive to initial conditions, and {\em a priori} unsuitable for set-based extensions).  Extensions of numerical methods of class 1 are thus given in \cite{MayneSR05,SchurmannA17b,SchurmannA17,SchurmannKA18},
while extensions of numerical methods of class 2 are given  in \cite{LeCoentF19,CoentF19,LygerosTS99,MitchellBT01,MitchellT03,ReissigR19}. These extensions have the respective advantages and disadvantages of their numerical counterparts. Set-based methods of class 1 are efficient (polynomial complexity in $n$-dimension of the problem, i.e., state vector dimension), but calculate {\em a priori} only {\em local} optimals. Set-based methods of class 2 compute global optimals, but undergo the
``curse of dimensionality'' (exponential complexity in the dimension $M$
of the state space), and are 
 limited to low dimensional problems.

Recently, in the numerical framework, Heymann et al.  \cite{Heymann18} have shown that, for certain problems, numerical methods of class~2 can give better solutions than numerical methods of class~1. They  have built a numerical solver, called ``Bocop'',
that implements both classes of methods %
\cite{Bocop}, and have given a set of examples that allows to evaluate 
and compare them \cite{BocopExamples}. 
We show in this paper that a set-based method of class~2 
combining DPP and a guaranteed Euler integration method \cite{LeCoentF19}, also allows us to 
compute 
approximate optimal solutions with good precision. Besides, these solutions enjoy the property of {\em robustness} against uncertainties on initial conditions and bounded disturbances.
We demonstrate the practical interest of our method 
on an example taken from the Bocop solver.
We also give a variant of our set-based method, inspired by the principle of Model Predictive Control \cite{Mayne14}, that allows us to compute 
approximate optimal solutions  more quickly, at the cost of losing the robustness property. 
\paragraph{Plan of the paper:}
In \cref{sec:robust}, we explain the principle of our method of optimal control synthesis, and give
the associated correctness results (convergence and robustness);
we compare the results of our method with those obtained 
by the Bocop numeric solver on an example of
Magnetic Resonance Imaging. In \cref{sec:receding},
we give an efficient variant of our method inspired by
the Model Predictive Control Approach
but observe the loss of the robustness property.
We conclude in \cref{sec:conclusion}.
\section{Robust optimal control}\label{sec:robust}
\subsection{Explicit Euler time integration}\label{ss:Euler}
We consider here  a time discretization of time-step $\tau$, and 
we suppose that the control law ${\bf u}(\cdot)$ is a {\em piecewise-constant} function, which takes its
values on a {\em finite} set $U$, called ``set of modes''.
Given $u\in U$, let us consider the differential system controlled by $u$: 
$$\frac{dy(t)}{dt}=f_u(y(t)).$$
where $f_u(y(t))$ stands for $f({\bf u}(t),y(t))$ with ${\bf u}(t)=u$ for $t\in[0,\tau]$.
We use $Y_{t,y_0}^u$ to denote  the exact continuous solution 
$y$ %
of the system at time~$t\in[0,\tau]$ under constant control $u$,
with initial condition~$y_0$.
This  solution is approximated using the 
{\em explicit Euler} integration method. We use $\tilde{Y}^u_{t,y_0}\equiv
y_0+tf_u(y_0)$ to denote Euler's approximate value of $Y^u_{t,y_0}$ for $t\in [0,\tau]$.

Given a sequence of modes (or ``pattern'') $\pi := u_1\cdots u_k\in U^k$, we denote by
$Y_{t,y_0}^{\pi}$
the solution of the system under  mode $u_1$ on $t\in [0,\tau)$
with initial condition~$y_0$,
extended continuously with the solution of the system under mode $u_{2}$ on $t\in[\tau,2\tau]$, and so on iteratively until mode $u_k$ on 
$t\in[(k-1)\tau,k\tau)$.
The control function ${\bf u}(\cdot)$ is thus piecewise constant  with ${\bf u}(t)=u_{n}$
for $t\in [(n-1)\tau,n\tau)$, $1\leq n\leq k$.
Likewise, we use
$\tilde{Y}_{t,y_0}^{\pi}$ to denote Euler's approximate value of $Y_{t,y_0}^\pi$
for $t\in [0,k\tau)$
defined by
$\tilde{Y}_{t,y_0}^{u_1\cdots u_n}=\tilde{Y}_{t,y_0}^{u_1\cdots u_{n-1}}+tf_{u_n}(\tilde{Y}_{t,y_0}^{u_1\cdots u_{n-1}})$ for $t\in [0,\tau)$ and $2\leq n\leq k$.
The approximate solution $\tilde{Y}_{t,y_0}^{\pi}$ is here a continuous piecewise linear function
on $[0,k\tau)$ starting at $y_0$.

\subsection{Finite horizon control problems}\label{ss:approx}

The optimization task is to find a control pattern $\pi\in U^k$ which guarantees that all states in a given set $S=[0,1]^M\subset \mathbb{R}^M$\footnote{We take here $S=[0,1]^M$ for the sake of notation simplicity, but $S$ can be any convex subset of $\mathbb{R}^M$.} are steered 
at time $t_{end}=k\tau$ as closely as possible to an end state $y_{end}\in S$.
Let us explain the principle of the method based on DPP and Euler integration
method used in~\cite{LeCoentF19,CoentF19}.
We consider the {\em cost function}: $J_{k}:S\times U^k\rightarrow \mathbb{R}_{\geq 0}$ 
defined by:
$$J_{k}(y,\pi)=\|Y_{k\tau,y}^{\pi}-y_{end}\|,$$
where 
$\|\cdot\|$ denotes the Euclidean norm in $\mathbb{R}^M$\footnote{We consider here the special case where the cost function is only made of a ``terminal'' subcost. The method extends to more general cost functions. Details will be given in the extended version of this paper.}.

We consider the {\em value function} ${\bf v}_k:S\rightarrow \mathbb{R}_{\geq 0}$
defined by:
$${\bf v}_k(y) := \min_{\pi\in U^k}\{J_{k}(y,\pi)\}\equiv
\min_{\pi\in U^k}\{\|Y_{k\tau,y}^{\pi}-y_{end}\|\}.$$ 

Given $k\in\mathbb{N}$ and $\tau\in\mathbb{R}_{>0}$, we consider the following {\em finite time horizon optimal control problem}: 
 Find for each $y\in S$
\begin{itemize}
\item the {\em value} 
${\bf v}_k(y)$, i.e.
$$\min_{\pi\in U^k}\{\|Y_{k\tau,y}^{\pi}-y_{end}\|\},$$

\item and an {\em optimal pattern}:
$$\pi_k(y) := arg\min_{\pi\in U^k}\{\|Y_{k\tau,y}^{\pi}-y_{end}\|\}.$$
\end{itemize}

In order to solve such optimal control problems, 
a classical ``direct'' method consists in
{\em spatially discretizing} the state space $S=[0,1]^M$
(i.e., the space of values of $y$).
We consider here a uniform partition of~$S$ into a finite number $N$
of cells of equal size:
in our case , this means that interval $[0,1]$ is divided into
$K$ subintervals of equal size, and $N=K^M$. A cell 
thus corresponds  to a $M$-tuple of subintervals. The center of a cell
coresponds to the $M$-tuple of the subinterval midpoints.
The associated grid ${\cal X}\subset S$ is the set of centers
of the cells of~$S$. The center~$z\in{\cal X}$ of a cell $C$ is considered as the $\varepsilon$-{\em representative} of 
all the points of $C$. We suppose that
the  cell size is such that $\|y - z\|\leq\varepsilon$,
for all $y\in C$ (i.e. $K\geq \sqrt{M}/2\varepsilon$).

We suppose that $S$ is ``controlled Euler-invariant'' in the sense that, for all
$y\in S$, there exits $u\in U$ such that %
$\tilde{Y}_{\tau,y}^u\in S$. We say that such $u$ is {\em admissible for $y$},
and we denote by $Adm(y)$ the (non-empty) set of modes admissible for $y\in S$.

In this context, the method proceeds as follows
(cf. \cite{LeCoentF19}): we consider the points of ${\cal X}$
as the vertices of a finite oriented graph;
there  is a connection from $ z\in {\cal X}$ 
to $ z'\in {\cal X}$ if $ z'$ is the $\varepsilon$-representative
of the Euler-based image $(z +\tau f_u( z))$ of~$z$, for some $u\in U$.
We then compute using dynamic programming
the ``path of length $k$ with minimal cost'' starting at $ z$: 
such a path is a sequence of $k+1$ 
connected points $ z\  z_k\  z_{k-1}\ \cdots\  z_1$ of
${\cal X}$ which minimizes the distance $\| z_1-y_{end}\|$.
This procedure allows us to compute
a pattern $\pi^\varepsilon_k(z)$ of length
$k$, which approximates the optimal pattern $\pi_k(y)$.

\begin{definition}
The function $next^{u}: {\cal X}\rightarrow {\cal X}$ 
is defined by:
\begin{itemize}
\item if $u\in Adm(z)$, then: %
    $next^{u}( z)= z'$, where
    $ z'\in{\cal X}\subset S$ 
    is the $\varepsilon$-representative of $\tilde{Y}_{\tau, z}^{u}$.
\item otherwise (i.e., $\tilde{X}_{\tau, z}^{u}\not\in  S$): $next^{u}(z)=\bot$.
\end{itemize}
\end{definition}
\begin{definition}
For all point $z\in {\cal X}$, the {\em spatially discrete
value function} ${\bf v}^{\varepsilon}_k:{\cal X}\rightarrow \mathbb{R}_{\geq 0}$ 
is defined by:
\begin{itemize}
\item for $k=0$, ${\bf v}_k^{\varepsilon}( z)=\| z-y_{end}\|$,
\item for $k\geq 1$,
    ${\bf v}_{k}^{\varepsilon}(z)=\min_{u\in Adm(z)}\{{\bf v}_{k-1}^{\varepsilon}(next^u( z))\}$.
\end{itemize}
\end{definition}
\begin{definition}
The {\em approximate  optimal pattern of length $k$}
associated
to $z\in{\cal X}$,
denoted by $\pi_k^{\varepsilon}( z)\in U^k$,  is 
defined by:
\begin{itemize}
\item if $k=0$, $\pi_k^{\varepsilon}( z)=\mbox{nil}$,
\item if $k\geq 1$, $\pi_k^{\varepsilon}( z) = {\bf u}_k( z) \cdot \pi'$
where
$${\bf u}_{k}( z)=arg \min_{u\in Adm(z)\subseteq U}\{{\bf v}_{k-1}^{\varepsilon}(next^u( z))\}$$
and $\pi' =\pi_{k-1}^{\varepsilon}( z')$
\ \ with \ $ z'=next^{{\bf u}_k( z)}( z)$.
\end{itemize}
\end{definition}
It is easy to
construct
a procedure $PROC_k^{\varepsilon}$ which takes a point $ z\in {\cal X}$ as input, and returns %
an approximate optimal pattern~$\pi_k^{\varepsilon}\in U^k$.

\begin{remark}
The complexity of  $PROC_k^\varepsilon$ is  
$O(m\times k\times N)$
where $m$ is the number of modes ($|U|=m$), 
$k$ the time-horizon length ($t_{end}=k\tau$)
and $N$ the number of cells of ${\cal X}$
($N=K^M$).
\end{remark}

\subsection{Correctness of the method}\label{ss:error}
Given a point $y\in S$ of $\varepsilon$-representative $z\in {\cal X}$,
and a pattern $\pi^\varepsilon_k$ returned by~$PROC_k^\varepsilon(z)$, we are now going to show that the distance
$\|\tilde{Y}_{k\tau,z}^{\pi^\varepsilon_k}-y_{end}\|$ converges to~${\bf v}_k(y)$
as $\varepsilon\rightarrow 0$.
We first consider the ODE:
$\frac{dy}{dt}=f_u(y)$, and give an upper bound to the error between
the exact solution of the ODE and
its Euler approximation (see~\cite{CoentF19,SNR17}).
\begin{definition}\label{def:delta}
Let $\mu$ be a given positive constant. Let us define, for all 
$u\in U$ and $t\in [0,\tau]$,
$\delta^u_{t,\mu}$ as follows:
$$\mbox{if } \lambda_u <0:\ \ 
\delta^u_{t,\mu}=\left(\mu^2 e^{\lambda_u t}+
 \frac{C_u^2}{\lambda_u^2}\left(t^2+\frac{2 t}{\lambda_u}+\frac{2}{\lambda_u^2}\left(1- e^{\lambda_u t} \right)\right)\right)^{\frac{1}{2}}$$
$$\mbox{if } \lambda_u = 0:\ \ 
\delta^u_{t,\mu}= \left( \mu^2 e^{t} + C_u^2 (- t^2 - 2t + 2 (e^t - 1)) \right)^{\frac{1}{2}}$$
$$\mbox{if } \lambda_u > 0:\ \ 
\delta^u_{t,\mu}=\left(\mu^2 e^{3\lambda_u t}+ 
\frac{C_u^2}{3\lambda_u^2}\left(-t^2-\frac{2t}{3\lambda_u}+\frac{2}{9\lambda_u^2}
\left(e^{3\lambda_u t}-1\right)\right)\right)^{\frac{1}{2}}$$
where $C_u$ and $\lambda_u$ are real constants specific to function $f_u$,
defined as follows:
$$C_u=\sup_{y\in S} L_u\|f_u(y)\|,$$
where $L_u$ denotes the Lipschitz constant for $f_u$, and
$\lambda_u$ is the OSL constant associated to $f_u$, i.e., the 
minimal constant such that, for all $y_1,y_2\in S$:
$$\langle f_u(y_1)-f_u(y_2), y_1-y_2\rangle \leq \lambda_u\|y_1-y_2\|^2,$$
where $\langle\cdot,\cdot\rangle$ denotes the scalar product of two vectors
of $S$.
\end{definition}

\begin{figure}[h!]
\centering
\includegraphics[scale=0.5]{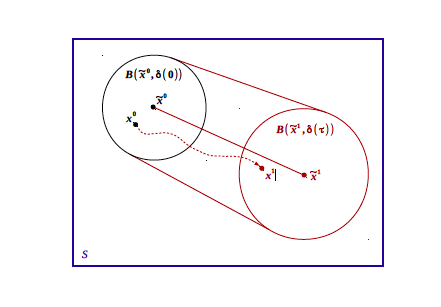}
\caption{Illustration of \cref{prop:basic}}
\label{fig:illustration}
\end{figure}

\begin{proposition}\label{prop:basic}\cite{SNR17}
Consider the solution $Y_{t,y_0}^u$ of $\frac{dy}{dt}=f_u(y)$ with
initial condition~$y_0$ of $\varepsilon$-representative $z_0$
(hence such that $\|y_0-z_0\|\leq\varepsilon$), 
and the approximate
solution $\tilde{Y}_{t,z_0}^u$ given by the explicit Euler scheme.
For all $t\in[0,\tau]$, we have:
$$\|Y_{t,y_0}^u-\tilde{Y}_{t,z_0}^u\|\leq \delta^u_{t,\varepsilon}.$$
\end{proposition}
\cref{prop:basic} underlies the principle of our set-based method
where set of points are represented as balls centered around the 
Euler approximate values of the solutions. This illustrated in~\cref{fig:illustration}: for any initial condition $x^0$ belonging
to the ball $B(\tilde{x}^0,\delta(0))$, the exact solution $x^1\equiv Y_{\tau,x^0}^u$ belongs to the ball $B(\tilde{x}^1,\delta(\tau))$ where $\tilde{x}^1 \equiv\tilde{Y}_{\tilde{x}^0,\tau}^u$ denotes the Euler approximation of the exact solution at 
$t=\tau$, and 
$\delta(\tau)\equiv\delta^u_{\tau,\delta(0)}$.

\begin{lemma}\label{lemma:1}\cite{CoentF19}
Consider the system $\frac{dy}{dt}=f_u(y)$
where the OSL constant $\lambda_{u}$ associated to $f_u$ is negative,
and initial error $e_0:=\|y_0-z_0\|>0$. 
Let
$G_u:=\frac{\sqrt{3}e_0|\lambda_u|}{C_u}$.
Consider the (smallest) positive root
$$\alpha_u := 1+|\lambda_{u}|G_{u}/4-\sqrt{1+(\lambda_{u}G_{u}/4)^2}$$
of equation:
$-\frac{1}{2}|\lambda_{u}|  G_{u}
+(2+\frac{1}{2}|\lambda_{u}|G_{u})\alpha-\alpha^2 = 0.$

Suppose:
$\frac{|\lambda_u|G_u}{4}<1.$%
Then we have $0<\alpha_u< 1$, and, for all $t\in[0,\tau]$
with $\tau\leq G_u(1-\alpha_u)$:
$$\delta_{e_0}^u(t) %
\leq e_0.$$ 
\end{lemma}
\begin{remark}\label{rk:subsampling}
If $\tau > G_u(1-\alpha_u)$, we can make use of {\em subsampling}, i.e.,
decompose~$\tau$ into a sequence of elementary time steps $\Delta t$
with $\Delta t \leq G_u(1-\alpha_u)$ in order to be still able to apply
\cref{lemma:1}. %
Let us point out that \cref{lemma:1} (and the use of subsampling) allows to ensure set-based reachability with the use of procedure $PROC_k^\varepsilon$. Indeed, in this setting, the explicit Euler scheme leads to decreasing errors, and thus, point based computations performed with the center of a cell can be applied to the entire cell. 
\end{remark}

We suppose henceforth that, for all $u\in U$, the system $\frac{dy}{dt}=f_u(y)$
satisfies:
$$(H):\ \ \ \lambda_u<0,\ \frac{|\lambda_u|G_u}{4}<1\  \mbox{ and }\  \tau \leq G_u(1-\alpha_u),\ \mbox{ for all } u\in U.$$
We have:

\begin{theorem}\label{th:2} (Convergence) \cite{CoentF19}.
Let $y\in S$ be a point
 of $\varepsilon$-representative $z\in {\cal X}$. Let
$\pi_k^\varepsilon$ be the pattern returned by $PROC_k^\varepsilon(z)$,
and $\pi^* := \mbox{argmin}_{\pi\in U_k} \|Y^\pi_{k\tau,y}-y_{f}\|$.
Let ${\bf v}_k(y) := \|Y_{k\tau,y}^{\pi^*}-y_{end}\|$ be the exact optimal value
of $y$. %
The approximate optimal value of~$y$,
$\|\tilde{Y}_{k\tau,y}^{\pi_k^\varepsilon}-y_{end}\|$, converges to 
${\bf v}_k(y)$ as $\varepsilon\rightarrow 0$.
\end{theorem}

\cref{th:2} formally justifies the correctness of our method
of optimal control synthesis by saying that 
the approximate optimal values computed by our method converge to the
exact optimal values when the mesh size tends to $0$.
Furthermore, we have:
\begin{theorem}\label{th:1} (Robustness)\cite{CoentF19}.
Let $y\in S$ be a point of of $\varepsilon$-representative
$z\in {\cal X}$,
and $\pi_k^\varepsilon$ the pattern returned by $PROC_k^\varepsilon(z)$.
We have:
$$\|Y_{t,y}^\pi-\tilde{Y}_{t,z}^\pi\|\leq \varepsilon,\ \ \ 
\mbox{ for all }\ \pi\in U^k \mbox{ and } t\in[0,k\tau].$$
It follows that, for two points $y_1,y_2\in S$ having the same $\varepsilon$-representative $z\in{\cal X}$,
we have:
$$|\|\tilde{Y}_{k\tau,y_1}^{\pi_k^\varepsilon}-y_{end}\|-\|\tilde{Y}_{k\tau,y_2}^{\pi_k^\varepsilon}-y_{end}\||\leq \varepsilon.$$
\end{theorem}

Last inequality of \cref{th:1} says that the approximate optimal values of $y_1$ and $y_2$
are equal up to $\varepsilon$.
This reflects the {\em robustness} of our method of optimal control synthesis against  {\em uncertainties  on  initial conditions}.
As for uncertainties  on initial conditions,
one has similar robustness results accounting for dynamical {\em bounded
disturbances}, as explained in \cref{appendixun}.

\subsection{Implementation}
The implementation of the robust and variant methods has been done in Python.
Each method corresponds to a program of
around 500 lines.
The source code is available at
	\href{https://lipn.univ-paris13.fr/~jerray/synchro/}{\nolinkurl{lipn.univ-paris13.fr/~jerray/synchro/}}.
In the experiments below, the program runs on a 2.80 GHz Intel Core i7-4810MQ CPU with 8\,GiB of memory.

\subsection{Example: Magnetic Resonance Imaging (MRI)}\label{ss:MRI}

Considering a system $q$ consisting of two different particles with spins $q_1, q_2$ (see \cite{BCCM13,BCCM14}).
The magnetization vectors $q_1 = (y_1, z_1) \in \mathbb{R}^2$ and $q_2 = (y_2, z_2) \in \mathbb{R}^2$ satisfy the differential system:

\begin{equation*}
q_1:
\begin{cases}
\overset{.}{y_1}  = 2 \pi T_{m} t_{m}(-\Gamma_1 y_1 - u_2 z_1)\\
\overset{.}{z_1} = 2 \pi T_{m} t_{m}(\gamma_1 (1 - z_1) + u_2 y_1)
\end{cases}
\label{q1}
\end{equation*}

\begin{equation*}
q_2:
\begin{cases}
\overset{.}{y_2}  = 2 \pi T_{m} t_{m} (-\Gamma_2 y_2 - u_2 z_2) \\
\overset{.}{z_2} = 2 \pi T_{m} t_{m} (\gamma_2 (1 - z_2) + u_2 y_2)
\end{cases}
\label{q2}
\end{equation*}
with: 
$\Gamma_1 = \frac{1}{T_{12}\Omega_{max}}$, 
$\gamma_1 = \frac{1}{T_{11}\Omega_{max}}$, 
$\Gamma_2 = \frac{1}{T_{22}\Omega_{max}}$, 
$\gamma_2 = \frac{1}{T_{21}\Omega_{max}}$, 
and $u_2 \in [-1, 1] $ the magnetic field (control).
Let $\Omega_{max}=202.95$, $T_{11}=2$, $T_{12}=0.3$, $T_{21}=2.5$, $T_{22}=2.5$, $T_{m}=26.17$ and $t_{m}=2$.
The goal is to make $q_1$ reach the origin $(0, 0)$ 
at a given time $t=t_{end}$ %
while maximizing 
the ``contrast'' $\|q_2(t_{end})-q_1(t_{end})\|=\|q_2(t_{end})\|$.
In order to account for the (soft) constraint $q_1(t_{end})=(0,0)$, we integrate
in the cost function $J_k$ a ``penalty term'' of the form
$\Vert q_1(t_{end})\Vert^2$. Our goal is thus to minimize
the terminal cost: $\alpha \Vert q_1(t_{end})^2\Vert - \beta \Vert q_2(t_{end})-q_1(t_{end}) \Vert^2$.
The domain $S$ of the states $(q_1,q_2)\equiv ((y_1,z_1),(y_2,z_2))$
is equal to $[-1,1]^2\times [-1,1]^2\equiv [-1,1]^4$.
The grid~${\cal X}$ corresponds to a discretization of
$S=[-1,1]^4$, where each component interval $[-1,1]$ is uniformly discretized into a set of $K$ points. 
The codomain $[-1,1]$ of the original continuous control function $u_2(\cdot)$ is itself discretized into a finite set~$U$ with our method. After discretization, $u_2(\cdot)$ is a 
piecewise-constant function that takes its values
in the finite set $U$ made of 30 values uniformly taken between $-1$ and $1$.
The function $u_2(\cdot)$ can change its value every $\tau$ seconds.
In the following experiments, we use the following parameter values:
$\alpha = 0.99$, $\beta = 0.01$, $\tau = 1/250$, $k=215$, $t_{end}=k\tau=0.86$, and $q_1(0)=(0,1)$. 
We will consider the cases $K=10$ (coarse grid) and $K=20$ (finer grid).
One can check that assumption $(H)$  is satisfied in both cases.
In order to test the robustness of the method, we will consider the cases $q_2(0)=(0,1)$ and $q_2(0)=(0.1,1)$.

\begin{figure}[h!]
\centering
\includegraphics[scale=0.35]{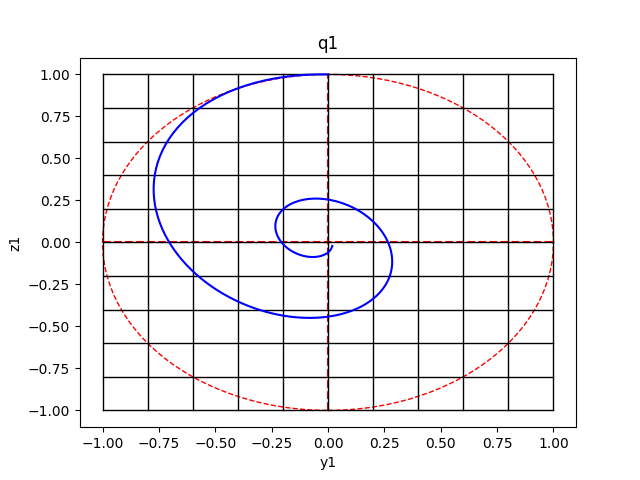}
\includegraphics[scale=0.35]{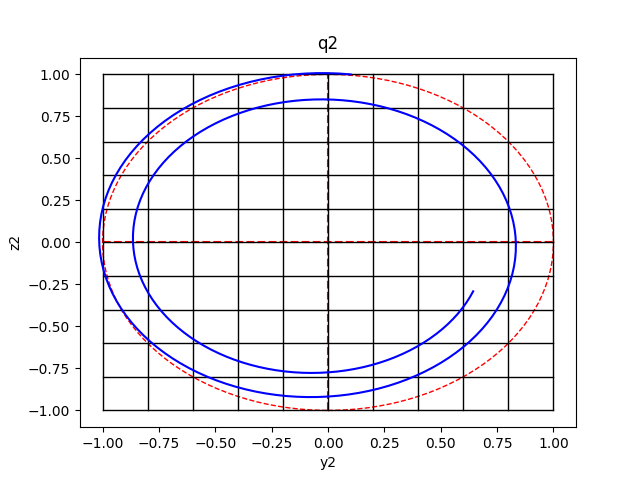}
\includegraphics[scale=0.35]{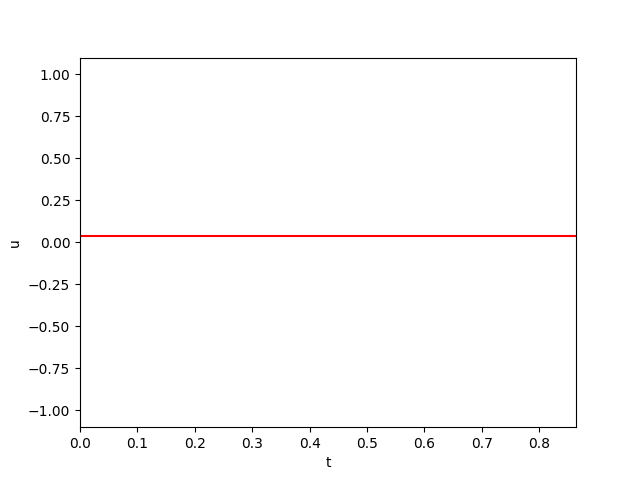}
\caption{Robust method applied to MRI for $K = 10$ and initial condition $q_2(0)=(0.1, 1)$, with $q_1=(y_1, z_1)$ (top left),  $q_2=(y_2, z_2)$ (top right) and control $u_2$ (bottom). When applied to $q_2(0)=(0,1)$, the method gives the same results.}
\label{fig:init00-grid10}
\end{figure}

For $K=10$
and $q_2(0)=(0, 1)$,  we have $q_2(t_{end}) = (0.6567, -0.2558)$, and
the optimal value of the contrast is $\|q_2(t_{end})\|= 0.7048$.
The CPU computation takes 389 seconds. See~\cref{fig:init00-grid10}.
For $q_2(0)=(0.1, 1)$, the synthesized control and the results 
are identical, which demonstrates the robustness of our method.

\begin{figure}[h!]
\centering
\includegraphics[scale=0.35]{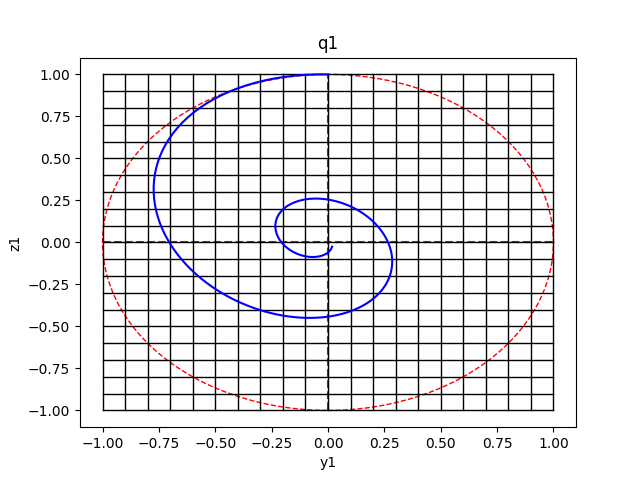}
\includegraphics[scale=0.35]{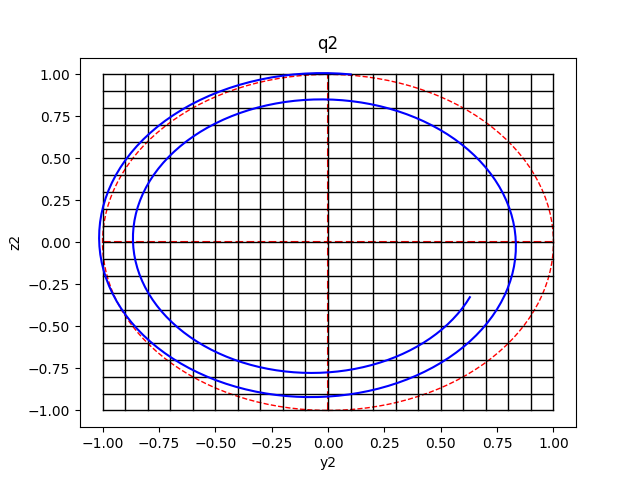}
\includegraphics[scale=0.35]{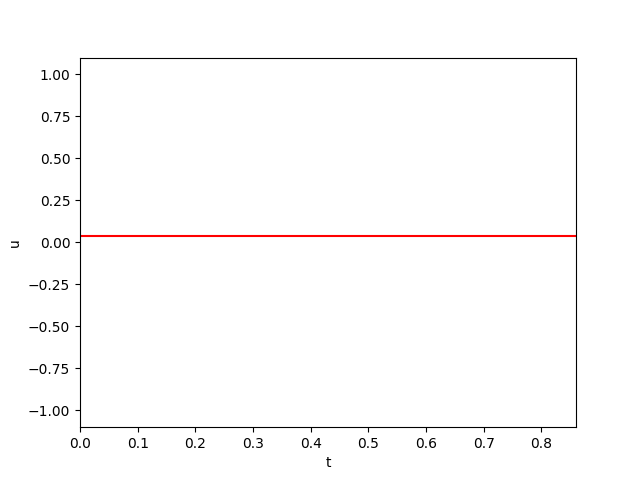}
\caption{Robust method applied to MRI for $K = 20$ and initial condition $q_2(0)=(0.1, 1)$, with $q_1=(y_1, z_1)$ (top left),  $q_2=(y_2, z_2)$ (top right) and control $u_2$ (bottom). When applied to $q_2(0)=(0,1)$, the method gives the same results.}
\label{fig:init00-grid20}
\end{figure}

For $K= 20$,
and $q_2(0)=(0, 1)$, we have 
$q_2(t_{end}) = (0.6439, -0.2913)$, and the contrast is
$\|q_2(t_{end})\|= 0.7067$.
(see \cref{fig:init00-grid20}). 
The CPU computation takes 3657 seconds.
See~\cref{fig:init00-grid20}.
For $q_2(0)=(0.1, 1)$, the synthesized control and the results 
are again identical, thus confirming the robustness of our method.

For comparison, we now perform the same experiments with the version of the numerical solver Bocop using convex optimization \cite{Bocop}.
For $q_2(0)=(0,1)$, we have with Bocop: $q_2(t_{end}) = (0.0499, -0.7938)$; the contrast is $\|q_2(t_{end})\|= 0.6746$.
The CPU computation time is 230 seconds.
See \cref{fig:init00-Bocop} (\cref{appendixdeux}).
For $q_2(0)=(0.1,1)$, we have, with Bocop: $q_2(t_{end}) = (0.0877,  -0.6631)$; the contrast is $\|q_2(t_{end})\|= 0.6689$.
The CPU computation time is 43 seconds.
See \cref{fig:init01-Bocop} (\cref{appendixdeux}).
We can see on this example that Bocop is {\em not} robust against slight changes
of initial conditions, the generated optimal trajectories being  very different from each other.
The optimal values of the contrast
computed by Bocop and our program are comparable.
However, the CPU times of Bocop are smaller
than those of our program (especially for $K=20$). 

\section{A Variant of the Method with Receding Horizon}\label{sec:receding}
The control computed by our method is robust, but its synthesis 
is time-costly because it requires a fine partition of the state space in order 
to decrease the error caused by the space discretization. 
We are now considering a variant
of our method, inspired by the Model Predictive Control Method (MPC) which uses a {\em receding horizon} \cite{Mayne14}.
In the original method, for a $k$-horizon problem ($t_{end}=k\tau$), to a point $y\in S$
is applied the optimal pattern $\pi(z)$ of length $k$ computed for the 
$\varepsilon$-representative $z$ of $y$ (returned by $PROC_k^\varepsilon(z)$).
In the variant inspired by MPC, we apply at point y only the first mode $u_1$
of $\pi(z)$, thus obtaining the point $y_1=y+\tau f_{u_1}(y)$.
Then, unlike the original method, we do not apply the second mode $u_2$
of $\pi(z)$, but we apply the first mode $u'_1$ of the optimal pattern 
$\pi(z_1$)
(returned by $PROC_k^\varepsilon(z_1)$),
where $z_1$ denotes the $\varepsilon$-representative of $y_1$. This
gives $y_2=y_1+\tau f_{u'_1}(y_1)$ (and not $y_1+\tau f_{u_2}(y_1)$ as before).
And so on, iteratively, one applies each time the first mode of the
optimal pattern  $\pi(z_n)$ returned by $PROC_k^\varepsilon(z_n)$,
where $z_n$ denotes the $\varepsilon$-representative of the solution~$y_n$ computed at $t=n\tau$ ($1\leq n\leq k-1$). 

\begin{figure}[h!]
\centering
\includegraphics[scale=0.35]{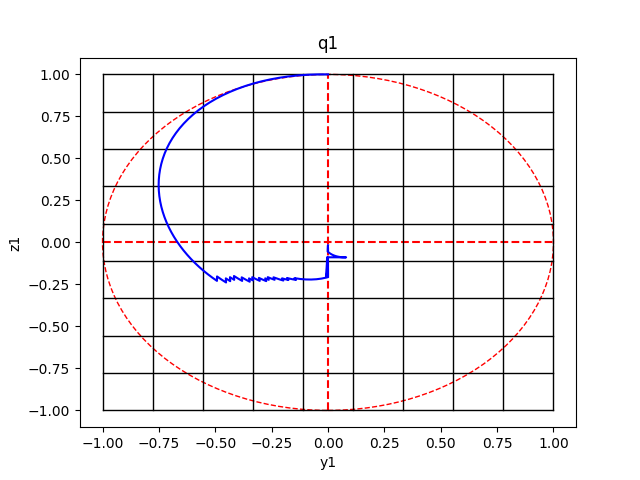}
\includegraphics[scale=0.35]{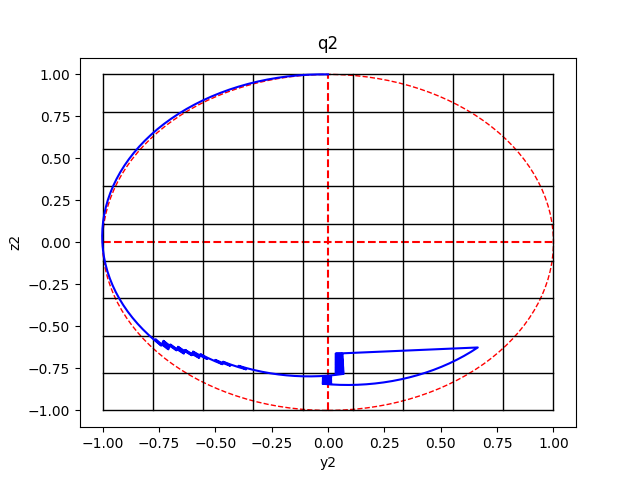}
\includegraphics[scale=0.35]{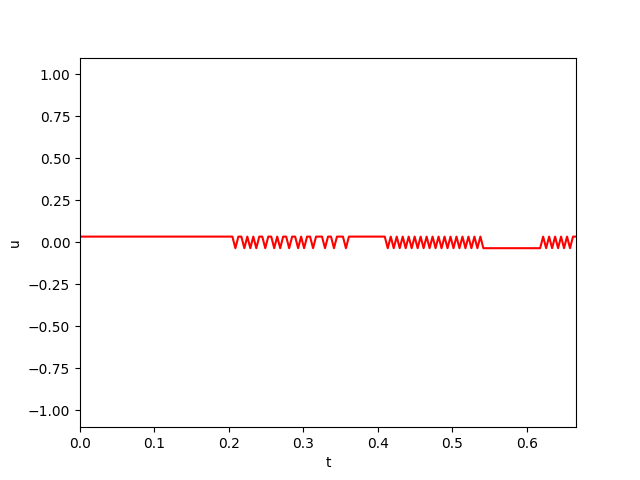}
\caption{Variant with receding horizon applied to MRI for initial condition $q_2(0)=(0, 1)$, with $q_1=(y_1, z_1)$ (top left),  $q_2=(y_2, z_2)$ (top right) and control $u_2$ (bottom).}
\label{fig:init00bis}
\end{figure}

\begin{figure}[h!]
\centering
\includegraphics[scale=0.35]{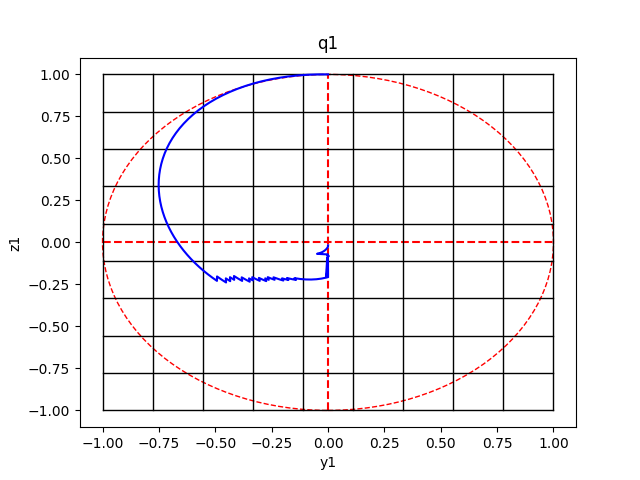}
\includegraphics[scale=0.35]{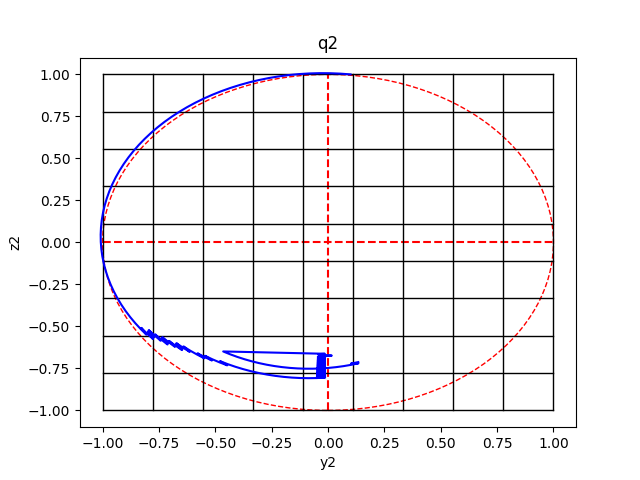}
\includegraphics[scale=0.35]{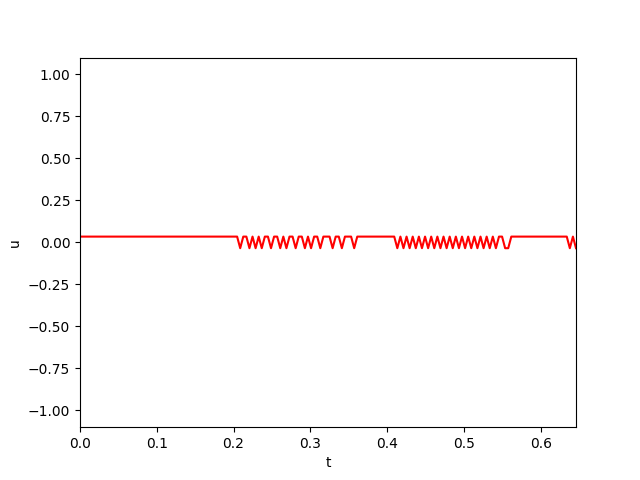}
\caption{Variant with receding horizon applied to MRI for initial condition $q_2(0)=(0.1, 1)$, with $q_1=(y_1, z_1)$ (top left),  $q_2=(y_2, z_2)$ (top right) and control $u_2$ (bottom).}
\label{fig:init01bis}
\end{figure}

This variant is not any longer
robust: trajectories from two close starting points do not usually stay close 
to each other anymore.
On the other hand, the computed values converge much faster to the
exact optimal values as $\varepsilon$ tends to~0. This allows us to
compute values of similar precision with the variant method, using a 
much coarser grid (bigger $\varepsilon$). 
The variant method is therefore more efficient than the original method.
We demonstrate this gain of efficiency and loss of robustness on the MRI example of \cref{ss:MRI}.
We first synthesize
the optimal control for $K=10$ and $q_2(0)=(0, 1)$, in which case 
we have: $q_2(t_{end}) = (0.0499, -0.7938)$, and the 
contrast is $\|q_2(t_{end})\|= 0.7954$.
(see \cref{fig:init00bis}). 
For $K=10$ and $q_2(0)=(0.1, 1)$, we have:
$q_2(t_{end}) = (0.1015, -0.7141)$, and the contrast
$\|q_2(t_{end})\|$ is $0.7210$.
(see \cref{fig:init01bis}). 
For both cases, the CPU computation takes 34 seconds.
We can see on this example that, unlike the original method, 
the variant method is {\em not} robust, a small difference between the initial conditions 
($q_2(0)=(0,1)$ {\em vs.} $q_2(0)=(0.1,1)$) leading to
very different trajectories.

For $K=20$ and $q_2(0)=(0,1)$,
we have $q_2(t_{end}) = (-0.06225, -0.5874)$, and the
contrast $\|q_2(t_{end})\|$ is  $0.5906$. (see \cref{fig:init00-grid20bis} in \cref{appendixtrois}).
The CPU computation now takes 443 seconds. 
For $K=20$ and $q_2(0)=(0.1,1)$, we have
$q_2(t_{end}) = (-0.1088, -0.7192)$,
and the contrast  $\|q_2(t_{end})\|$ is $0.7274$.
(see \cref{fig:init01-grid20bis} in \cref{appendixtrois}).
The CPU computation now takes 501 seconds.

On the MRI example, the CPU times of the variant method are thus 
much smaller than those of the original method, and comparable 
to those of Bocop. %
Besides, the optimal values of the contrast
computed by the variant method are
slightly better than those computed by Bocop.
The variant method is thus more efficient than the original method,
but does not retain its robustness property.
There is therefore a trade-off to be found between robustness (guaranteed 
with the original method)
and efficiency (obtained with the MPC variant).

The results of \cref{sec:robust,sec:receding} for $x_2(0)=(0.1,1)$ are recapitulated
in \cref{tab:recap}.
\begin{table}[htp]
	\begin{centering}
		\begin{tabular}{|l|c|c|c|c|c|}
		\hline
			& robust method(K=10) & robust method(K=20) & variant(K=10) & variant(K=20) & Bocop\\ 
		\hline
		Robust?   &     yes        &       yes        &   no        &       no      &  no     \\
		\hline		
		Contrast: & 0.7048    	  &     0.7669        &   0.7210     &     0.7273        & 0.6746      \\
		\hline		
		CPU time (s): & 389            &     3657         &    34         &   501         & 230       \\
		\hline
		\end{tabular}
		\caption{Summary table of results}
		\label{tab:recap}
	\end{centering}
\end{table}

\section{Conclusion}\label{sec:conclusion}

As pointed out in \cite{Heymann18,Bocop}, 
numerical methods of optimal control, based on DPP, can  
compete on low dimensional
examples,
with methods based on convex optimization. 
Along these lines, we show in this paper that 
a set-based method of optimal control 
combining DPP and a guaranteed Euler integration method, allows to 
synthesize a correct-by-design optimal
control that is {\em robust} against uncertainties on initial conditions
and bounded disturbances. 
We have demonstrated the practical interest of our method on an example
taken from the numerical Bocop solver. We have observed similar 
results in experiments with other case studies from Bocop, that will be given in the 
extended version of this paper. We have also
considered a variant of our method with a receding horizon, that makes the
control synthesis more efficient at the cost of losing the robustness property.
There is therefore a trade-off to be found between robustness (guaranteed 
with the original method)
and efficiency (obtained with the variant using a receding horizon).

\newcommand{\LNCS}{LNCS}

\ifdefined\VersionAuthor
	\renewcommand*{\bibfont}{\small}
	\printbibliography[title={References}]
\else
	\bibliographystyle{splncs04} %
	\bibliography{rp20}
\fi

\newpage
\appendix
\section{Robustness against bounded disturbances}\label{appendixun}
A differential system with ``bounded disturbances'' is of the form 
$$\frac{dy(t)}{dt}=f_u(y(t),w(t)),$$
with $u\in U$, $t\in [0,\tau]$,
states $y(t)\in\mathbb{R}^M$, and disturbances $w(t)\in{\cal W}\subset \mathbb{R}^d$ (${\cal W}$ is compact, i.e., closed and bounded).
See, e.g., \cite{SchurmannA17b}. We assume that any possible disturbance trajectory is bounded in the compact set
${\cal W}$ for $t\in [0,\tau]$.
We use $\phi_{u}(t;y^0,w(\cdot))$ to denote the solution of
$\frac{dy(t)}{dt}=f_u(y(t),w(t))$ for $t\in[0,\tau]$ with $y(0)=y^0$.
If we consider an undisturbed system, we use
$\phi_u(t; y^0,0)$ 
(resp. $\tilde{\phi}_u(t; y^0,0)$ )
to denote the solution (resp. the approximate Euler solution) without disturbances, i.e.,
${\cal W}=0$.

Given a pattern $\pi=u_k\cdots u_1\in U^k$, these notations extend naturally to $t\in [0,k\tau]$ by considering the solutions
obtained by applying successive modes $u_k,\dots, u_1$ in a continuous manner.
The optimization task is now to find a control pattern $\pi\in U^k$ which guarantees that all states in $S\subset \mathbb{R}^M$ are steered 
at time $t=k\tau$ as closely as possible to an end state $y_{end}$, {\em despite the
 disturbance set} ${\cal W}$. 

We now suppose that $S$ is controlled Euler-invariant for the
{\em undisturbed} system, i.e.: for all $y\in S$,
there exists $u$ such that $\tilde{\phi}_u(\tau;y,0)\in S$.
We also suppose (see \cite{AdrienRP17}) that, for all $u\in U$,
there exist constants $\lambda_u\in\mathbb{R}_{<0}$ 
and $\gamma_u\in\mathbb{R}_{\geq 0}$ such that,
for all $y_1,y_2\in S$ and $w_1,w_2\in {\cal W}$:
$$\langle f_u(y_1,w_1)-f_u(y_2,w_2), y_1-y_2\rangle \leq 
\lambda_u\|y_1-y_2\|^2 + \gamma_u \|y_1-y_2\|\|w_1-w_2\|\ \ \ \ \ (H1).$$

We now give a version of \cref{prop:basic} with bounded disturbance $w(\cdot)\in{\cal W}$.
\begin{proposition}\label{prop:1bis}\cite{AdrienRP17}
Given
a sampled switched system with bounded disturbance 
of the form $\{\frac{dy(t)}{dt}=f_u(y(t),w(t))\}_{u\in U}$
satisfying (H1) for all $u\in U$, consider a point~$y_0\in S$ of $\varepsilon$-representative $z^0\in {\cal X}$.
We have,
for all $w(\cdot) \in {\cal W}$,  $u\in U$: %

$$\|\phi_{u}(\tau;y^0, w(\tau))-\tilde{\phi}_u(\tau;z^0,0)\|\leq \delta^u_{\tau,\varepsilon,{\cal W}}.$$
with %
\begin{multline}
 \delta^u_{t,\varepsilon,{\cal W}} = 
 \left( \frac{C_{u}^2}{-\lambda_{u}^4} \left( - \lambda_{u}^2 t^2 - 2 \lambda_{u} t + 2 e^{\lambda_{u} t} - 2 \right) \right.   \\
 + \left. \frac{1}{\lambda_{u}^2} \left( \frac{C_{u} \gamma_{u} |{\cal W}|}{-\lambda_{u}} \left( - \lambda_{u} t + e^{\lambda_{u} t} -1 \right) \right. \right.  \\ + \left. \left. \lambda_{u} \left( \frac{\gamma_{u}^2 (|{\cal W} |/2)^2}{-\lambda_{u}} (e^{\lambda_{u} t } - 1) + \lambda_{u} \varepsilon^2 e^{\lambda_{u} t}  \right) \right)  \right)^{1/2}
\end{multline}

\end{proposition}

\cref{th:2,th:1} can themselves be extended 
to account for bounded disturbance $w(\cdot)\in{\cal W}$. The details will be given 
in the extended version of this paper.

\newpage
\section{Sensitivity of Bocop to Initial Conditions}\label{appendixdeux}
\begin{figure}[h!]
\centering
\includegraphics[scale=0.30]{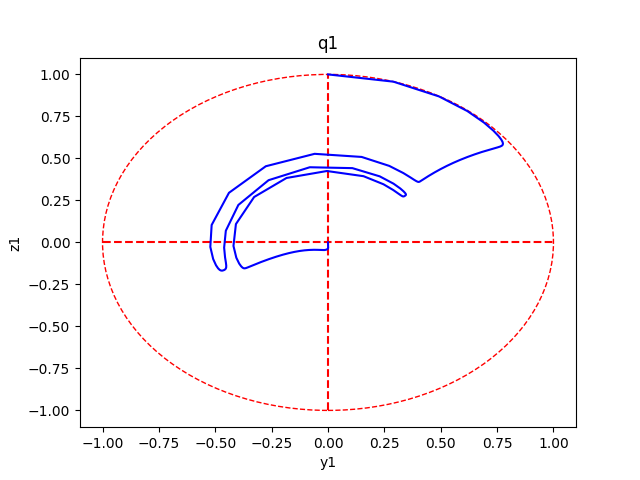}
\includegraphics[scale=0.30]{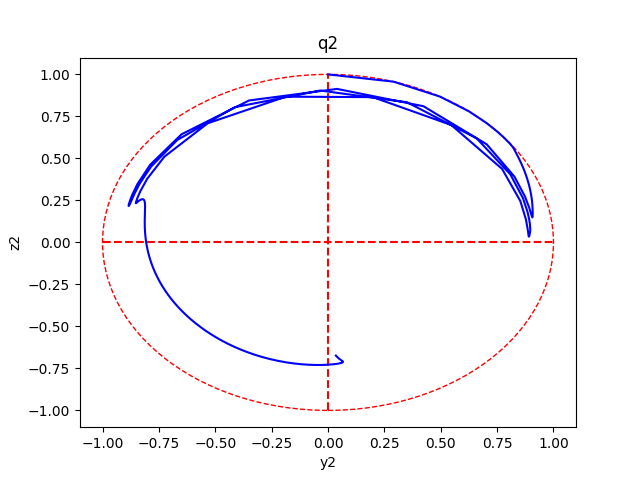}
\includegraphics[scale=0.25]{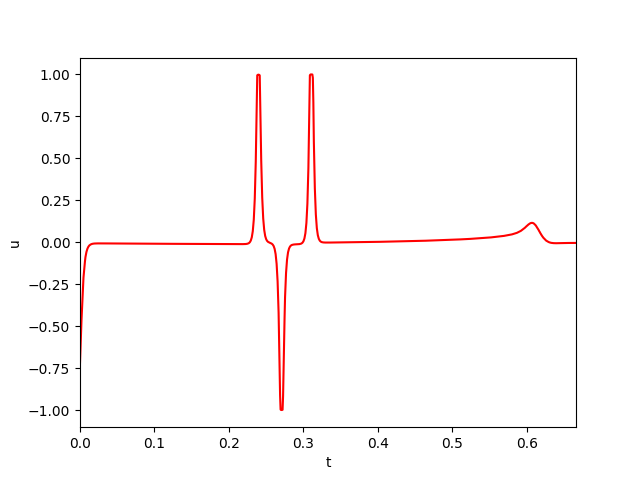}
\caption{Bocop solution on MRI when initially $q_2(0)=(0, 1)$, with $q_1=(y_1, z_1)$ (top left), $q_2=(y_2, z_2)$ (top right) and control $u_2$ (bottom).}
\label{fig:init00-Bocop}
\end{figure}
\begin{figure}[h!]
\centering
\includegraphics[scale=0.30]{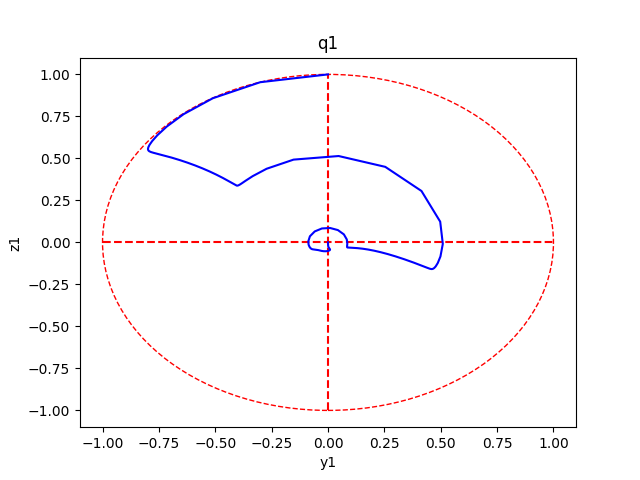}
\includegraphics[scale=0.30]{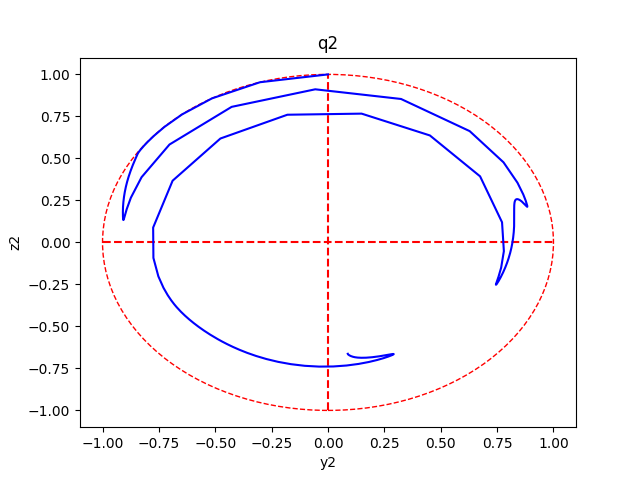}
\includegraphics[scale=0.25]{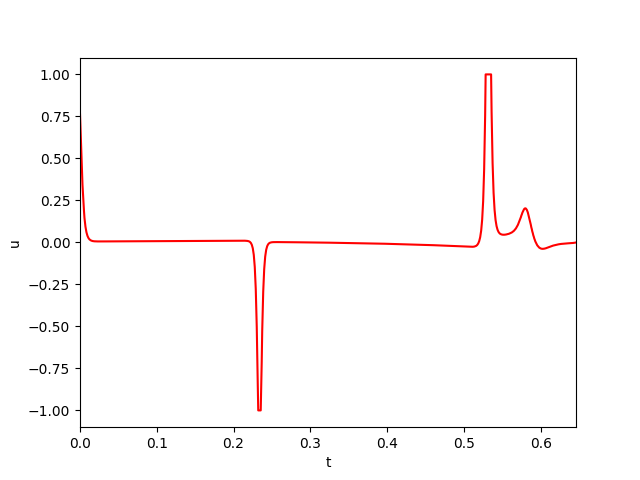}
\caption{Bocop solution on MRI for initial condition $q_2(0)=(0.1, 1)$, with $q_1=(y_1, z_1)$ (top left),  $q_2=(y_2, z_2)$ (top right) and control $u_2$ (bottom).}
\label{fig:init01-Bocop}
\end{figure}

\newpage
\section{Results of variant with receding horizon applied to MRI with $K=20$}\label{appendixtrois}

\begin{figure}[h!]
\centering
\includegraphics[scale=0.35]{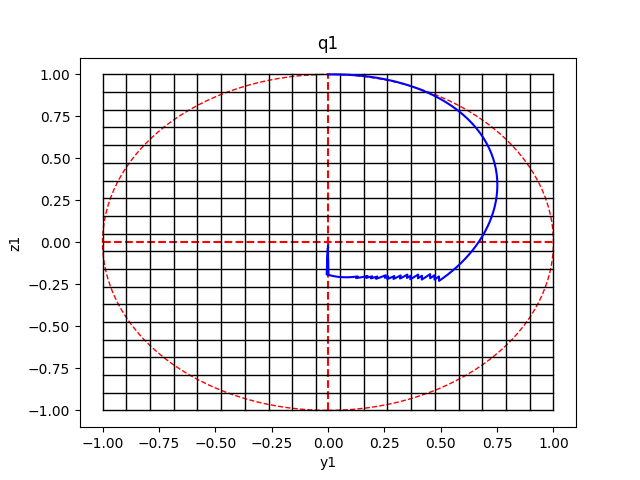}
\includegraphics[scale=0.35]{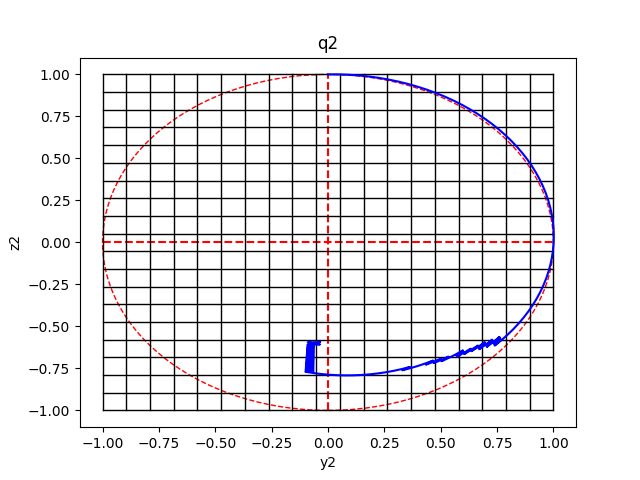}
\includegraphics[scale=0.35]{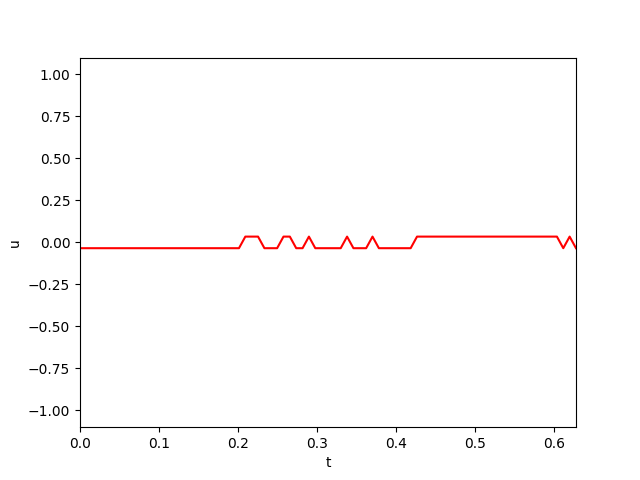}
\caption{Variant with receding horizon applied to MRI with a finer grid ($K=20$) for initial condition $q_2(0)=(0, 1)$, with $q_1=(y_1, z_1)$ (top left),  $q_2=(y_2, z_2)$ (top right) and control $u_2$ (bottom).}
\label{fig:init00-grid20bis}
\end{figure}

\begin{figure}[h!]
\centering
\includegraphics[scale=0.35]{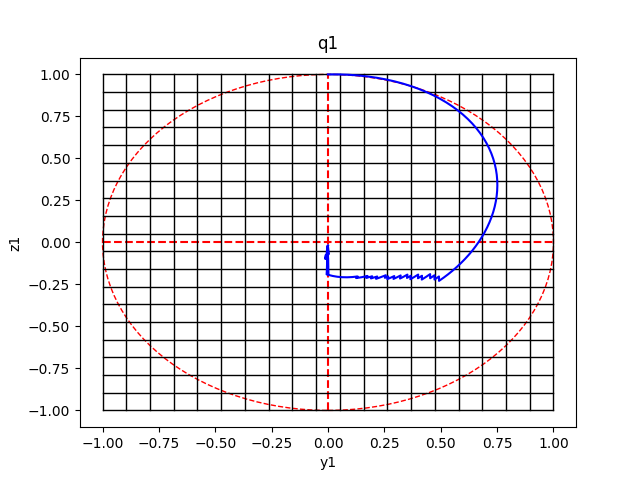}
\includegraphics[scale=0.35]{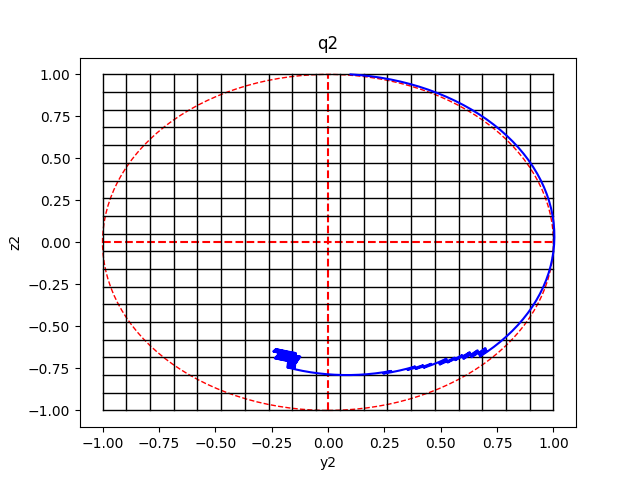}
\includegraphics[scale=0.35]{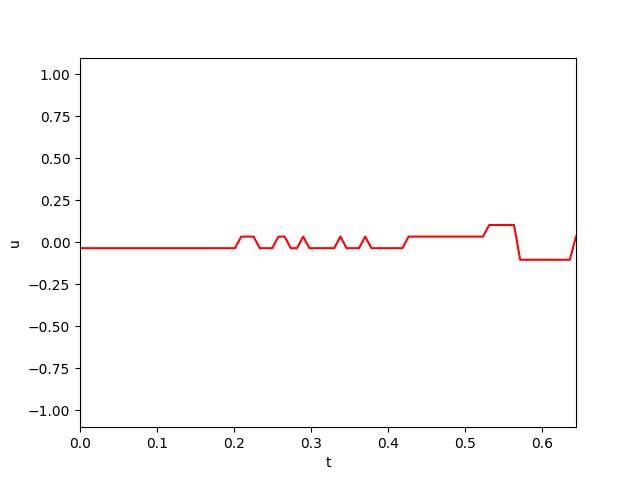}
\caption{Variant with receding horizon applied to MRI with a finer grid ($K=20$) for initial condition $q_2(0)=(0.1, 1)$, with $q_1=(y_1, z_1)$ (top left),  $q_2=(y_2, z_2)$ (top right) and control $u_2$ (bottom).}
\label{fig:init01-grid20bis}
\end{figure}
\end{document}